\documentclass[preprint,aps,prd,nofootinbib,showpacs,12pt,epsfig]{revtex4-1}

\usepackage{mathrsfs,amsmath,amssymb,amsfonts}
\usepackage{graphicx,graphics,epsfig,hhline,color}

\usepackage{graphicx}
\usepackage{amssymb}
\usepackage{amsmath}
\usepackage{pifont}
\usepackage{epsfig}
\usepackage[utf8]{inputenc}

\newcommand{\be}{\begin{eqnarray}}
\newcommand{\ee}{\end{eqnarray}}
\newcommand{\beq}{\begin{eqnarray}}
\newcommand{\eeq}{\end{eqnarray}}

\newcommand{\Z}{\mathbb{Z}}

\newcommand{\expm}{e^{-\beta\left(E_i-\mu\right)}}
\newcommand{\expmm}{e^{-2\beta\left(E_i-\mu\right)}}
\newcommand{\expmmm}{e^{-3\beta\left(E_i-\mu\right)}}
\newcommand{\expp}{e^{-\beta\left(E_i+\mu\right)}}
\newcommand{\expppp}{e^{-3\beta\left(E_i+\mu\right)}}

\newcommand{\bc}{\begin{center}}
\newcommand{\ec}{\end{center}}

\newcommand{\dfp}{\frac{d^4p}{(2\pi)4}}

\newcommand{\delsl}{\partial\hspace{-1.95mm}/}

\newcommand{\psl}{p\hspace{-1.5mm}/}

\newcommand{\unity}{1\hspace{-1.3mm}1}
\newcommand{\ave}[1]{\langle{#1}\rangle}

\newcommand{\tr}[1]{{\rm Tr}\,[{#1}]}

\begin{document}
\title{Interplay between chiral and axial symmetries in   a SU(2)
Nambu--Jona-Lasinio Model with the Polyakov loop
}
\author{M. C. Ruivo}
\altaffiliation[Corresponding author]{M. C. Ruivo}
\email{maria@teor.fis.uc.pt}
%
\author{M. Santos}
\email{mario@teor.fis.uc.pt}
%
\author{P. Costa}
\email{pcosta@teor.fis.uc.pt}
%
\author{C. A. de Sousa}
\email{celia@teor.fis.uc.pt}
\affiliation{Departamento de F\'{\i}sica, Universidade de Coimbra,
P-3004-516 Coimbra, Portugal, EU}
\date{\today}

\begin{abstract}
We consider a two flavor Polyakov--Nambu--Jona-Lasinio (PNJL) model  where the Lagrangian
includes an interaction term that explicitly breaks the U$_A(1)$ anomaly.
At finite temperature, the restoration of chiral and axial symmetries, signaled by the
behavior of several observables, is investigated.
We compare the effects of two regularizations at finite temperature, one of them, that allows
high momentum quarks states, leading to the full recovery of chiral symmetry.
From the analysis of the behavior of the topological susceptibility and of the mesonic
masses of the  axial partners, it is found in the SU(2) model that, unlike the SU(3) results,
the recovery of the axial symmetry is not a consequence of the full recovery of the chiral symmetry.
Thus, one needs to use an additional idea, by means of a temperature dependence of the
anomaly coefficient, that simulates instanton suppression effects.
\end{abstract}
\maketitle


\section{Introduction}
\label{sec:introduction}

Spontaneous breaking and restoration of chiral symmetry is one of the most fundamental
aspects inferred from quantum chromodynamic theory (QCD) and is manifested through the
equation of state of hot and dense hadronic matter. The Nambu--Jona-Lasinio (NJL) model
is one of the most studied effective models of QCD at finite temperature and has been
applied successfully to reproduce the low-energy phenomena, although  this model does not
support some features of QCD, such as quark confinement.

With the intention of broadening the applicability of effective QCD models, in
particular, allowing for the implementation of additional features of the QCD
phenomenology, the Polyakov Loop is considered within the NJL model, that so is endowed
with a mechanism that allows the analysis of the expected confinement/deconfinement phase
transition in the hadron-quark system. This extended model, the so called
Polyakov--Nambu--Jona-Lasinio (PNJL) model, first implemented in Ref.
\cite{Meisinger:1996PLB}, provides a simple framework which considers the chiral and the
confinement order parameters. In fact, the NJL model describes the interactions between
constituent quarks, hence providing the correct chiral properties. The static gluonic
degrees of freedom,  introduced in the NJL Lagrangian through an effective gluon
potential in terms of the Polyakov loop, take into account features of deconfinement
\cite{Meisinger:1996PLB,Fukushima:2004PLB,Ratti:2005PRD,Megias:2006PRD}. A good approach
to reproduce lattice results is given by the coupling of the quarks to the Polyakov loop.
This leads to a reduction weight of the quark degrees of freedom as the critical
temperature is approached from above (interpreted as a manifestation of confinement).

Besides the chiral symmetry, another important symmetry, explicitly broken in the QCD
Lagrangian, is the axial U$_A$(1) symmetry~\cite{hooft}, which might also be restored at
extreme temperatures. In fact, there are indications from the lattice calculations that,
at high temperatures, effects arising from the U$_A$(1) breaking are strongly suppressed.
This suggests an effective restoration of the U$_A$(1) symmetry. An important issue that
the present paper intends to analyze is the possible restoration of axial and chiral
symmetries, which observables may give indications regarding this restoration, and
whether there is an interplay between both restorations or not. By this we will use the
PNJL model in the SU(2) sector but with the inclusion of a term on the Lagrangian that
explicitly breaks the U$_A$(1) symmetry. It is worth noting that the effects of the
anomaly have been intensively studied in the NJL model in SU(3) with the 't Hooft
term~\cite{kle18,hatk21,prsy,prsy1,ruivo}, and more recently, the possibility of axial
symmetry restoration and its effects on diverse observables \cite{chi1,prsy,prsy1,ruivo}.
The model in SU(2), due to its simplicity, allows to isolate some aspects of the problem,
reason why its study can lead to a relevant contribution for the understanding of physics
associated to the breaking and restoration of the U$_A$(1) symmetry. Interest in studying
this issue in two flavor models is manifested in recent works \cite{gatto}.

The same model has been used in Ref. \cite{Brauner:2009gu} which particularly looks into
the nature of the superfluid phase. Here we will focus on the topological susceptibility,
the chiral and axial partners and the comparison between the results of SU(2) and SU(3)
models.

We will investigate the behavior of the following observables with temperature: quark
condensates, topological susceptibility and  $\pi$, $\sigma$, $a_0$ and $\eta$ mesons.
Two approaches will be followed in order to understand the mechanism of restoration of
axial symmetry: we will consider different degrees of U$_A$(1) symmetry breaking in the
vacuum, by using different values for the strength of the anomaly that are kept constant
with varying temperature; alternatively, we  fix the degree of symmetry breaking in the
vacuum and allow the coupling strength to decrease with temperature. We will also
consider the effects  upon the observables behavior of using two types of regularization:

\textit{- Regularization I}: as the cutoff $\Lambda$ is only necessary to regularize some
integrals in the vacuum  and it is not necessary at finite $T$, it  is considered
infinite as soon as thermal effects are considered ($\Lambda \rightarrow \infty$). This
has the effect of allowing for  high quark moments;

\textit{- Regularization II}: the cutoff is taken always with a finite value, at zero or finite
$T$ ($\Lambda\, =\, const.$),  and all the momenta couple with the same strength up to a cutoff
momentum $\Lambda \sim 1$~GeV.

In  Sec. II we will introduce the formalism   for a two flavor  NJL type model with two
different interacting parts that allow to disentangle chiral and axial symmetries. The
gap equations and the mesonic propagators will also be  discussed, as well as the
breaking of the axial symmetry and the topological susceptibility. The Polyakov loop
extension of the model will be explained as well. The different types of regularization
will be presented in the context of the model extension for finite temperature. We
redirect the details of the calculations to the appendices included in a previous
paper~\cite{prsy}. In Sec. III we present and discuss our results for the phase
transition and also for the temperature behavior of the meson masses and the topological
susceptibility, aiming at discussing  restoration of  symmetries. We conclude in Sec. IV
with a brief summary.


\section{Formalism}
\label{sec:formalism}

The aim of this section is to present  the mathematical formalism of the NJL model and
its extension to the PNJL model. In the present work, a  two flavor, three color, quark
system will be studied, and it will be explicitly introduced a term on the Lagrangian
that acts as a source of anomaly. The anomaly of the SU(2) model is already present in
the original NJL model~\cite{NJL,NJL2} through a Fierz transformation and has been
considered in other works as well (see Ref. \cite{buballa2} and Refs. there in), the
strength of the anomaly coupling being the same as the chiral coupling. In the present
work we have an extra term that ensures an independent mechanism of axial symmetry
breaking.

\subsection{A SU(2) Nambu--Jona-Lasino type model}
\label{subsec:NJLA}

The Lagrangian that will be used here is a SU(2) version of the NJL model taking into
account an additional term that, although being a chiral invariant, explicitly
breaks the axial symmetry.

We start with a NJL type model with two flavors defined by  the following Lagrangian:
\begin{eqnarray}
\mathcal{L} &=& \bar{q}(i \delsl - m)q + \mathcal{L}_1 + \mathcal{L}_2,\label{e1_}
\end{eqnarray}
with two different interacting parts
\begin{eqnarray}
\mathcal{L}_1 &=& g_1 \Big{[}(\overline{q} q)^2 + (\bar{q} i \gamma _5 \vec{\tau} q)^2 +
(\bar{q} \vec{\tau} q)^2 + (\bar{q} i \gamma _5 q)^2 \Big{],}
\label{e1_la}
\end{eqnarray}
\begin{eqnarray}
\mathcal{L}_2 &=& g_2 \Big{[}(\bar{q} q)^2 + (\bar{q} i \gamma _5 \vec{\tau} q)^2 -
(\bar{q} \vec{\tau} q)^2 - (\bar{q} i \gamma _5 q)^2 \Big{].}
\label{e1_lb}
\end{eqnarray}

The quark fields $q = (u,d)$ are defined in Dirac and color fields, respectively
with two flavors, $N_f=2$ and three colors, $N_c=3$, the coupling constants   $g_1$ and
$g_2$ have  dimension $energy^{-2}$, and  $\hat{m}=\mbox{diag}(m_u,m_d)$ is the current
quark mass matrix.

The \label{e1} Lagrangian is  chiral invariant in the limit where the current quark
masses vanish. Both terms $\mathcal{L}_1$ and $\mathcal{L}_2$  are invariant upon
SU(2)$_{L}\otimes$SU(2)${_R}\otimes$U(1) type transformations, but the $\mathcal{L}_2$
component makes the Lagrangian non-covariant upon U$_A$(1) transformations. The
$\mathcal{L}_2$ term, that may be represented in the form of a determinant:
\begin{eqnarray}
\mathcal{L}_2 &=& 2 g_2 \Big{[} \mbox{det} \big{[}\bar{q} (1 + \gamma_5) q \big{]}+
\mbox{det} \big{[}\bar{q} (1 - \gamma_5) q \big{]}\Big{]} ,
\label{e01_lb}
\end{eqnarray}
can be identified as an interaction induced by instantons,
according to 't Hooft, and it explicitly breaks the axial symmetry even in the chiral limit.

In this model, while the vector current is conserved, the axial isovector current is only
conserved in the chiral  limit and, as a consequence of the $\mathcal{L}_2$  term in the
Lagrangian, the isoscalar axial current is not conserved even in the chiral limit. In fact:
\begin{eqnarray}
\partial_{\mu} j^{\mu}_5 = 2 m   (\bar{q}\ i \gamma_5 q)
+ 8 g_2 \Big{[} (\bar{q}  \vec{\tau} q) (\bar{q} i \gamma _5 \vec{\tau} q)
-(\bar{q} q)(\bar{q} i \gamma _5 q)\Big{].}
\label{e02_lb}
\end{eqnarray}
%
The last equation is the equivalent of the QCD 4-divergence of $j^{\mu}_5$:
\begin{eqnarray}
 \partial_{\mu} j^{\mu}_5 = 2 m   (\bar{q}\ i \gamma_5 q) + 2 N_F Q(x),
 \label{e03_lb}
\end{eqnarray}
which allows to identify the topological charge $Q(x)$ in our model as:
\begin{eqnarray}
 Q(x)= 2 g_2  \Big{[} (\bar{q}  \vec{\tau} q) (\bar{q} i \gamma _5 \vec{\tau} q)-(\bar{q} q)(\bar{q} i
 \gamma _5 q)\Big{]},
 \label{e04_lb}
\end{eqnarray}
or, equivalently, as
\begin{eqnarray}
Q(x)= 2 g_2  \Big{[} \mbox{det} \big{[}(\bar{q} (1 - \gamma_5) q - \mbox{det} \big{[}(\bar{q} (1 + \gamma_5)
\Big{]}.
\end{eqnarray}

This equation shows the important role played by the coupling constant $g_2$ in our
analysis bearing in mind that the physical effects of the U$_A(1)$ anomaly are only
manifested  with non-zero topological charge.

Having identified the topological charge, the topological susceptibility may be
calculated \cite{chi1}, as it will be shown in Sec. II.C.

The four fields in the original Lagrangian  can be rearranged in the form:
\begin{eqnarray}
\mathcal{L}_1 + \mathcal{L}_2 = \frac{g_s}{2} \Big{[}(\bar{q} q)^2 + (\bar{q} i \gamma _5
\vec{\tau} q)^2 \Big{]} + \frac{g_a}{2}\Big{[}(\bar{q} \vec{\tau} q)^2 + (\bar{q} i \gamma _5 q)^2
\Big{]},
\label{e2}
\end{eqnarray}
where $g_s=2\,({g_1+g_2})$ and $g_a=\, 2\,({g_1-g_2})$.

From the last expression it is  easy to  understand the reason why the Lagrangian
supports four mesonic channels: ${\cal O}_\sigma = \unity$, ${\cal O}_{{\pi}} = i\gamma_5
{\vec{\tau}}$, ${\cal O}_\eta= i\gamma_5$ and ${\cal O}_{\vec{a_{0}}} = \unity
{\vec{\tau}}$. After integrating the generating functional over quark fields and using
$\sigma,{\vec{\pi}},\eta,   {\vec{a_{0}}}$ as auxiliary fields, the following effective
action is obtained:
\begin{eqnarray}
I_{eff} &=& -i \, \text{Tr }\, \text{ln} \Big{(} -i \delsl - m +\sigma + i \gamma _5 {\vec {\tau}} {\vec{\pi}}+
i \gamma _5 \eta +{\vec {\tau}} {\vec{a_{0}}}  \Big{)} - \frac{\sigma^2 + \vec{\pi}^2}{2g_s}-\frac{\eta^2 + \vec{a_0}^2}{2g_a}.
\label{e12}
\end{eqnarray}

A standard calculation leads straightforwardly to the  gap equation and to the meson
propagators. The interaction term  of the Lagrangian with  $g_s$ coupling is responsible
for the calculation of the propagators of the chiral partners (${{\pi}}\,, \sigma$),
while the term associated to $g_a$ allows for the chiral partners ($\eta\,, {{a_0}}$).
Since we consider equal quark masses, flavor mixing effects induced by the axial symmetry
breaking are not visible.

The following gap equations are  obtained:
\begin{equation}
M_{i}=m_{i}-2g_{_{S}}\left\langle\bar{q}q \right\rangle_{i},
\label{gap2}
\end{equation}
where one identifies $i=u,d$ and $M_{i}$ as the constituent quark mass. The quark
condensates are determined by
\begin{equation}
\left\langle\bar{q}q \right\rangle_{i}=-i\mbox{Tr}\frac{1}{\hat{p}-M_{i}}=-i\mbox{Tr}
S_{i}(p),
\end{equation}
being  $S_i(p) = (\psl - M_i+i\varepsilon)^{-1}$ the propagator of quarks.

The mass spectra of the mesons is obtained by the analysis of the pole structure of the
meson propagator, given by
\begin{equation}
   1 - 2 g_{s,a} \,\Pi_{\cal M}(q^2=M_{\cal M}^2) = 0.
\label{rpamass}
\end{equation}
where
\begin{equation}
   \Pi_{\cal M}(q2) \;=\; i \int \dfp \tr{{\cal O}_{\cal M}\,S(p+q)\,{\cal O}_{\cal M}\,S(p)} \label{pol}
\end{equation}
is the polarization operator for the quark-antiquark system regarding  the channel with
quantum numbers $\{{\cal M}\}$ in the mesonic sector. As mentioned above, $g_{s}$  is
related  to $\pi$ and $\sigma$ mesons and $g_{a}$ to $\eta$ and $a_0$ mesons.

\subsection{Extension to the PNJL model}

Now we include the  Polyakov loop and its effective potential to the NJL type model described above.
The Lagrangian of this SU(2)$\otimes$SU(2) quark model with explicit chiral symmetry
breaking where the quarks couple to a (spatially constant) temporal background gauge
field (represented in term of Polyakov loops) is given by \cite{Pisa1,Ratti:2005PRD}:
\begin{eqnarray}
{\mathcal L_{PNJL}\,}&=& \bar q\,(\,i\, {\gamma}^{\mu}\,D_\mu\,-\,\hat m)\,q \nonumber
+ \frac{g_s}{2}\, [(\bar{q} q)^2 + (\bar{q} i \gamma _5 \vec{\tau} q)^2 ] +
\frac{g_a}{2}\,[(\bar{q} \vec{\tau} q)^2 + (\bar{q} i \gamma _5 q)^2 ]\\
&-& \mathcal{U}\left(\Phi[A],\bar\Phi[A];T\right).
\label{eq:lag}
\end{eqnarray}

%
The quarks are coupled to the gauge sector {\it via} the covariant
derivative $D^{\mu}=\partial^\mu-i A^\mu$. The strong coupling
constant $g_{Strong}$ has been absorbed in the definition of $A^\mu$:
$A^\mu(x) = g_{Strong} {\cal A}^\mu_a(x)\frac{\lambda_a}{2}$ where
${\cal A}^\mu_a$ is the SU$_c(3)$ gauge field and $\lambda_a$ are the
Gell--Mann matrices. Besides in the Polyakov gauge and at finite temperature
$A^\mu = \delta^{\mu}_{0}A^0 = - i \delta^{\mu}_{4}A^4$. \\
The Polyakov loop $\Phi$ (the order parameter of $\Z_3$ symmetric/broken phase transition
in pure gauge) is the trace of the Polyakov line defined by:
$ \Phi = \frac 1 {N_c} {\langle\langle \mathcal{P}\exp i\int_{0}^{\beta}d\tau\,
A_4\left(\vec{x},\tau\right)\ \rangle\rangle}_\beta$.

The pure gauge sector is described by an effective potential
$\mathcal{U}\left(\Phi[A],\bar\Phi[A];T\right)$ chosen to reproduce at
the mean-field level the results obtained in lattice calculations:
\begin{equation}
    \frac{\mathcal{U}\left(\Phi,\bar\Phi;T\right)}{T^4}
    =-\frac{a\left(T\right)}{2}\bar\Phi \Phi +
    b(T)\mbox{ln}[1-6\bar\Phi \Phi+4(\bar\Phi^3+ \Phi^3)-3(\bar\Phi \Phi)^2],
    \label{Ueff}
\end{equation}
where
\begin{equation}
    a\left(T\right)=a_0+a_1\left(\frac{T_0}{T}\right)+a_2\left(\frac{T_0}{T}
  \right)^2\,\mbox{ and }\,\,b(T)=b_3\left(\frac{T_0}{T}\right)^3\, .
\end{equation}

The effective potential exhibits the feature of a phase transition from color confinement
($T<T_0$, { the minimum of the effective potential being at $\Phi=0$}) to color
deconfinement ($T>T_0$, the minima of the effective potential occurring at $\Phi \neq
0$).

The parameters of the effective potential $\mathcal{U}$ are given in Table \ref{table:param}.
These parameters have been fixed in order to reproduce the lattice
data for the expectation value of the Polyakov loop and QCD thermodynamics in the pure
gauge sector \cite{Kaczmarek:2002mc,Kaczmarek:2007pb}.

\begin{table}[t]
    \begin{center}
        \begin{tabular}{cccc}
            \hhline{|----|}
            $a_0$ & $a_1$ & $a_2$ & $b_3$ \\
            \hline
            3.51  & -2.47 & 15.2 & -1.75  \\
            \hhline{|----|}
        \end{tabular}
         \caption{\label{table:param}
         Parameters for the effective potential in the pure gauge sector.}
    \end{center}
\end{table}

The parameter $T_0$ is  the critical temperature for the deconfinement phase transition
within a pure gauge approach: it was fixed to $270$ MeV, according to lattice findings.
This choice ensures an almost exact coincidence between chiral crossover and deconfinement at
zero chemical potential, as observed in lattice calculations.

The PNJL grand canonical  potential density in the SU$_f$(2) sector can be written as
\cite{Ratti:2005PRD,Hansen:2007PRD}:
\begin{widetext}
\be
\Omega(\Phi, \bar\Phi, M ; T, \mu)
&=&{\cal U}\left(\Phi,\bar{\Phi},T\right)
+2g_{_{S}} N_f\left\langle\bar{q_{i}}q_{i}\right\rangle^2
- 2 N_c\,N_f \int_\Lambda\frac{\mathrm{d}^3p}{\left(2\pi\right)3}\,{E_p}\nonumber \\
&-& 2N_f\,T\int\,\frac{\mathrm{d}^3p}{\left(2\pi\right)^3}\,\left( z^+_\Phi(E_i) +
z^-_\Phi(E_i) \right) , \label{omega} 
\ee
\end{widetext}
where  $E_i$ is the quasi-particle energy for the quark $i$:
$E_{i}=\sqrt{\mathbf{p}^{2}+M_{i}^{2}}$, and  $z^+_\Phi$ and $z^-_\Phi$ are the partition
function densities.

The explicit expression of $z^+_\Phi$ and $z^-_\Phi$ are given by:
\begin{align}
z^+_\Phi(E_i)
&\equiv& \mathrm{Tr}_c\ln\left[1+ L^\dagger \expp\right]=\ln\left\{1+3\left(\bar\Phi
+\Phi \expp \right)\expp+\expppp \right\}, \label{eq:termo1}
\\
z^-_\Phi(E_i) &\equiv& \mathrm{Tr}_c\ln\left[1+ L \expm \right]= \ln\left\{ 1 + 3\left(
\Phi + \bar\Phi \expm \right) \expm
 + \expmmm \right\}.
\label{eq:termo2}
\end{align}
A word is in order to describe the role of the Polyakov loop in the present model. Almost
all physical consequences of the coupling of quarks to the background gauge field stem
from the fact that in the expression of $z_\Phi$,  $\Phi$ or $\bar\Phi$ appear only as a
factor of the one- or two-quarks (or antiquarks) Boltzmann factor, for example $\expm$
and $\expmm$. Hence when $\Phi,\bar\Phi \rightarrow 0$ (signaling what we designate as
the ``confined phase'') only $\expmmm$ remains in the expression of the grand canonical
potential, leading to a thermal bath with a small quark density. At the contrary
$\Phi,\bar\Phi \rightarrow 1$ (in the ``deconfined phase'') gives a thermal bath with all
1-, 2- and 3-particle contributions and a significant quark density.

This formalism, presented here for completeness in the grand canonical approach, will be
employed in the present work with $\mu=0$. This condition implies $\Phi=\bar\Phi=0$.

\subsection{The topological susceptibility}
\label{subsec:topsuc}

The topological susceptibility, $\chi$, is an essential parameter for the study  of the
breaking and restoration of the U$_A$(1) symmetry.

The topological susceptibility is defined as:
\begin{equation}
 \chi=\int{\rm d}^4x\; \langle 0 |TQ(x)Q(0)|0 \rangle_{\rm c},
\end{equation}
where $c$ means connected diagrams and $T$ the time order operator.
Using the definition of Eq. (\ref{e04_lb}) it  can be seen that the operator $Q(x) Q(0)$ in this model reads:
\begin{eqnarray}
Q(x)Q(0) &=& 4 g_2^2 \Big{[} [\bar{q}(x) q(x)][\bar{q}(x) \gamma_5 q(x)]-[\bar{q}(x) \vec{\tau }q(x)][\bar{q}(x) \gamma_5 \vec{\tau} q(x)]\Big{]}  \nonumber \\
&& \Big{[} [\bar{q}(0) q(0)][\bar{q}(0) \gamma_5 q(0)]-[\bar{q}(0) \vec{\tau} q(0)][\bar{q}(0) \gamma_5 \vec{\tau} q(0)]\Big{]}  \nonumber\\
&=& 4 (G \alpha)^2 \Big{[} [\bar{q}(x) q(x)][\bar{q}(x) \gamma_5 q(x)][\bar{q}(0) q(0)][\bar{q}(0) \gamma_5 q(0)] \nonumber\\
&-&  [\bar{q}(x) q(x)][\bar{q}(x) \gamma_5 q(x)][\bar{q}(0) \vec{\tau} q(0)][\bar{q}(0) \gamma_5 \vec{\tau} q(0)]  \nonumber\\
&-&  [\bar{q}(x) \vec{\tau }q(x)][\bar{q}(x) \gamma_5 \vec{\tau} q(x)] [\bar{q}(0) q(0)][\bar{q}(0) \gamma_5 q(0)] \nonumber\\
&+&  [\bar{q}(x) \vec{\tau }q(x)][\bar{q}(x) \gamma_5 \vec{\tau} q(x)] [\bar{q}(0)
\vec{\tau} q(0)][\bar{q}(0) \gamma_5 \vec{\tau} q(0)] \Big{]}. \label{asus1}
\end{eqnarray}

\begin{figure}[t]
\begin{center}
  \begin{tabular}{cc}
        \includegraphics[width=0.5\textwidth]{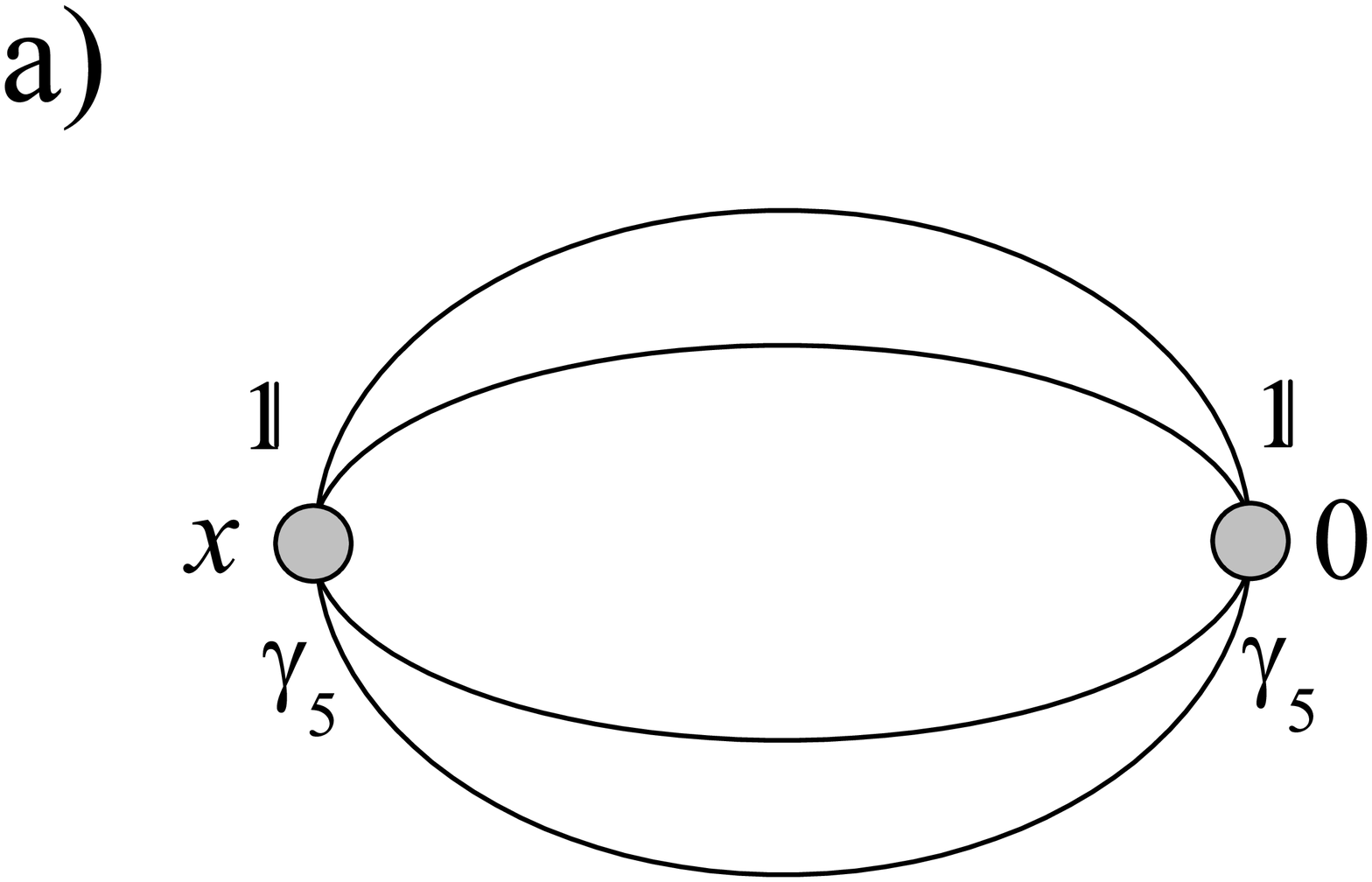} &
        \includegraphics[width=0.5\textwidth]{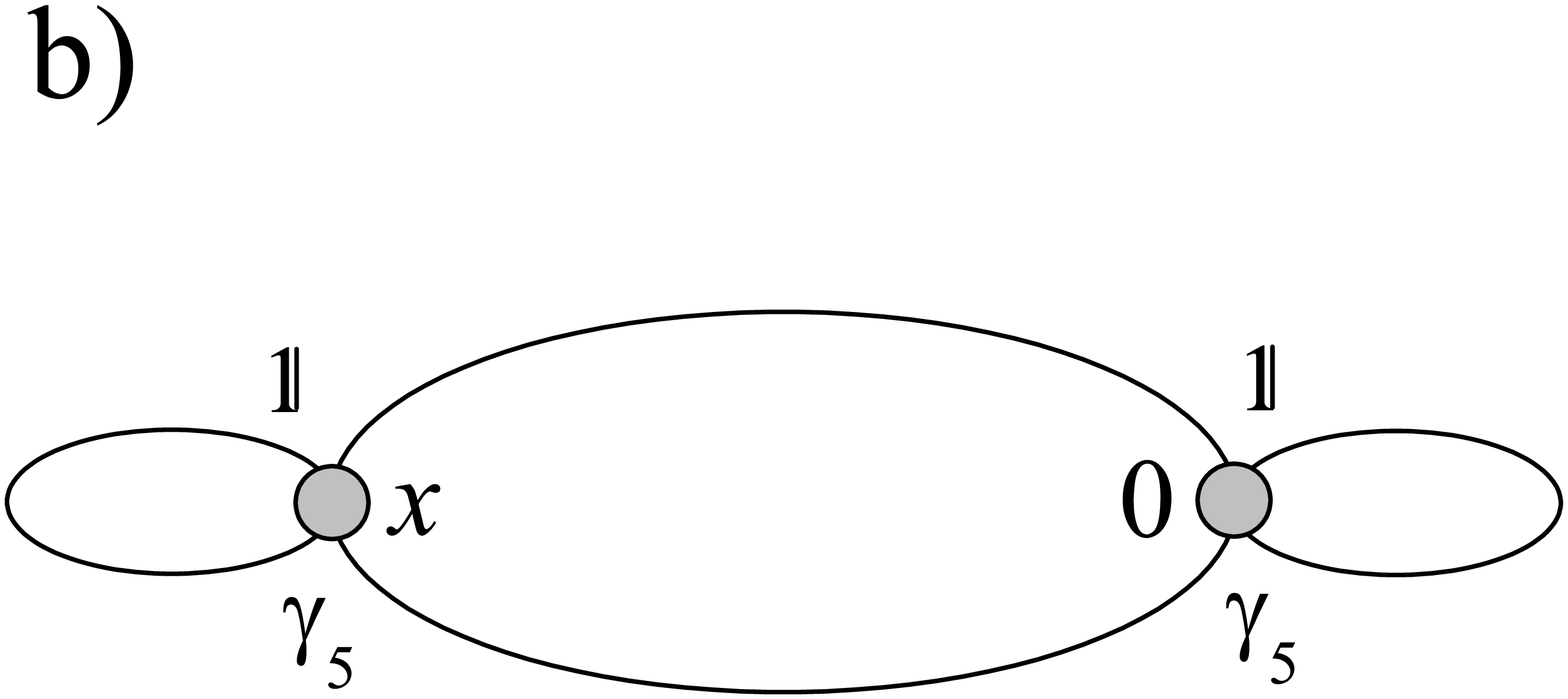}\\
        \hspace*{-0.25cm}\includegraphics[width=0.5\textwidth]{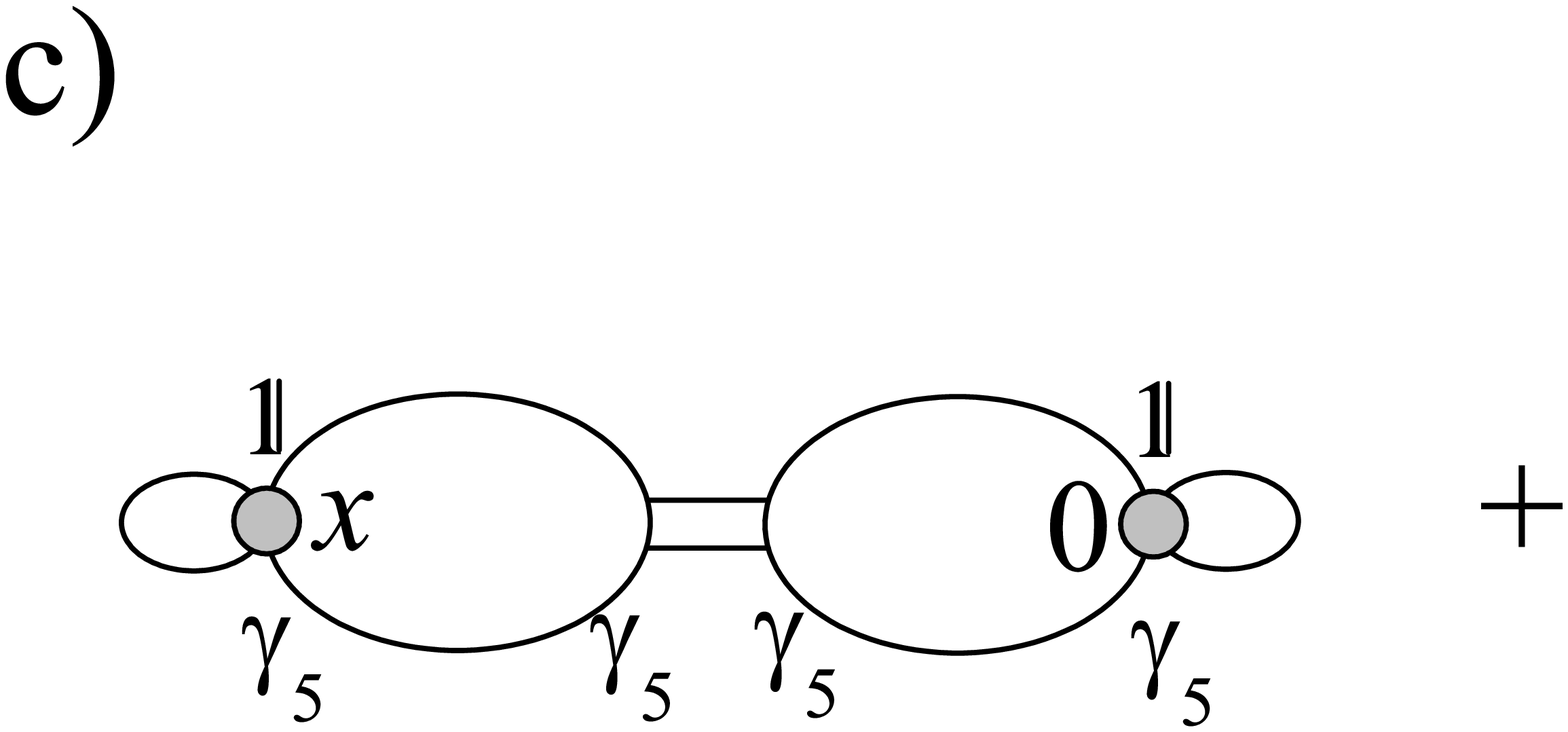} &
        \hspace*{-0.5cm}\includegraphics[width=0.5\textwidth]{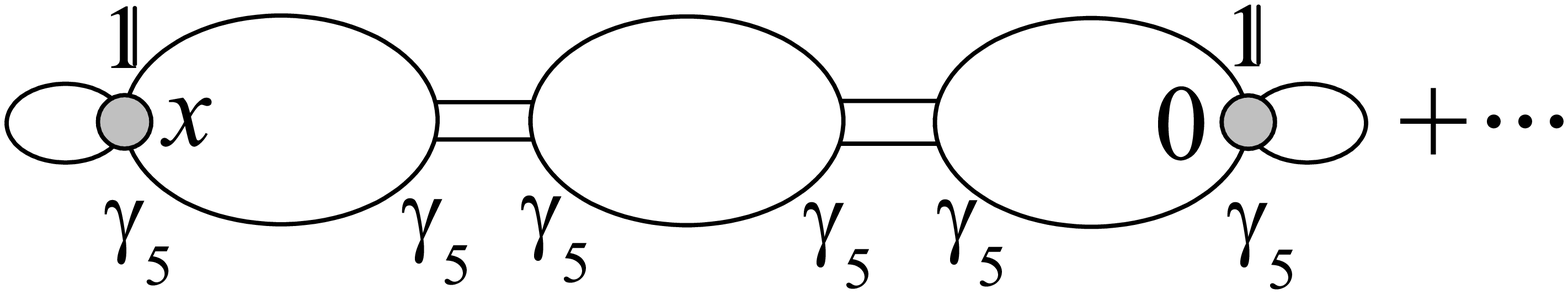}
    \end{tabular}
\end{center}
\caption{Connected diagrams for the loop ((a) and (b)) and for the ring contributions (c).
Only (b) and (c) are of order $1/N_c$.}
\label{Fig:diag}
\end{figure}

Taking  into account only the connected diagrams of order $1/N_c$ (see Fig.
\ref{Fig:diag}) and  following a similar approach to \cite{chi1},  we arrive at the
following expression for the topological susceptibility:
\begin{equation}
 \chi = 4 N_f\,  g_2^2\,\langle \bar q q \rangle^2\, \frac{ 4 I_1} {1- 8 g_a I_1}.
\label{chi4}
\end{equation}

As already mentioned, one of the aims of this paper is to perform a comparison between
the behavior with temperature of the topological susceptibility in the framework of the
SU(2) and SU(3) PNJL models with explicit axial symmetry breaking. As we will see, some
meaningful differences will appear, so it is useful, in order to understand these
differences, to compare both expressions. For the sake of simplicity we will write the
SU(3) expression for the case of equal  quark masses  $m_u=m_d=m_s$:

\begin{equation}
\chi = 4 N_f\,  \left(g_D \langle \bar q q \rangle\right)^2\,\langle \bar q q \rangle^2\,
\frac{ 4 I_1} {1- 8 P_{00} I_1} \label{chi42},
\end{equation}
%
where $P_{00}= (g_S - 2 g_D \left\langle  \bar q q\right\rangle)$
(for details concerning the SU(3) model see the Appendix \ref{appen} and Ref. \cite{Costa:2009}).
A comparison shows that the  expressions for quantities with effects of  the anomaly in
SU(2) depend on a constant, $g_2 $, while  the corresponding expressions in SU(3) depend
on a constant multiplied by a quark condensate,  $g_D   \left\langle  \bar q q\right\rangle$.
The same  replacement holds  for the  gap equations and for the propagators of
the chiral partners ($\eta, a_0$).

We will return to this point again when discussing the
behavior of several observables with temperature where it will be shown that some effects are
different in SU(2) and SU(3) due to the fact that in the last model the ``effective''
constant ${g_D}_{eff}= g_D \langle \bar q q \rangle$ already depends on temperature
through the quark condensate, while its corresponding in SU(2), $g_2$, remains
constant, unless we explicitly allow  a variation with
temperature.


\section{Results}

\subsection{Vacuum properties and parameters fixing}

\label{subsec:results}

The present PNJL model has  four parameters in the NJL sector: $m$, $\Lambda$, $g_1$ and $g_2$.
We choose to adjust the parameters in vacuum by fitting to well known experimental data or lattice values:
the mass of the pion, its decay constant, the quark condensate and the topological susceptibility,
$\chi$, that  are shown in Table \ref{table:param2}.
The masses of the $\sigma$, $\eta$ and $a_0$  mesons come as outputs.

Our main purpose is to study the behavior of several observables with temperature, in
particular those that might signal the chiral and deconfinement phase transitions, and
the effective restoration of chiral and axial symmetries. For this we will use two
different approaches, that  will be presented  now in the vacuum, and its consequences
for the behavior with temperature will be explored in the next subsection. The two
scenario to be considered are:

{\it Scenario A} -  We will keep $g_1$ and $g_2$ as independent parameters.
At finite temperature we may allow $g_2$ to be temperature dependent but $g_1$ is kept constant.
This is equivalent to the usual  treatment of  the SU(3) model;

{\it Scenario B} - We redefine the coupling constants such as the   set ($g_1, g_2$) will
be replaced by ($ G,\, \alpha$) in the following parametrization:
\begin{equation}
\label{redef}
g_1= G\, (1\,-\, \alpha), \,\,\,\,\,\,\,\,\, g_2= G\, \alpha,
\end{equation}
with  $\alpha \,\epsilon \,\{0,1\} $ \cite{buballa2,Brauner:2009gu}. In this picture
$g_1$ and $g_2$ are not   independent, but  $g_s=2(g_1 + g_2)= 2 G$ will be  kept  always
constant; on the contrary, $g_a= 2 G (1-2 \alpha)$ varies with $\alpha$.

Consequently, when studying the effect of the degree of axial symmetry breaking, by choosing different
values for $\alpha$, only the topological susceptibility and the masses of $\eta, a_0$ will be affected,
since they depend explicitly on $\alpha$; the $\pi$ and $\sigma$ channels will not be affected.


\begin{table}[t]
\begin{center}
\begin{tabular}{||c||c|c|c|c|c|c|c||}
\hline\hline
& $f_\pi$ & $\ave{\bar qq}^{1/3}$  & $m_\pi$ & $m_\sigma$ & $m_\eta$ & $m_{a_0}$ & $\chi^{1/4}$\\
& [MeV]   &  [MeV]                 & [MeV]   & [MeV]      & [MeV]    & [MeV]     & [MeV]\\
\hline
\hline
Model & 93 & -300 & 140.2 & 803.7  &  704.5 & 919.8 & 180.8  \\
\hline
Exp. /Latt. & 92.4   &  -270   &  135.0 & 400-1200 & 547.3 & 984.7 & 180 \\
\hline\hline
\end{tabular}
\end{center}
\caption{Numerical values for the calculated observables: $f_\pi$, $\ave{\bar qq}$, the
meson masses and the topological susceptibility, obtained with $\Lambda= 590$ MeV, $G
\Lambda^2 = 2.435$, $\alpha=0.2$ and $m=6$ MeV.} \label{table:param2}
\end{table}

Let us now discuss {\it Scenario B} for the vacuum state considering special  values for
the parameter $\alpha$:
\begin{itemize}
  \item [(1)] $\alpha=0$: the coupling constant $g_2=0$ and  the  axial symmetry is unbroken.
  \item [(2)] $\alpha=1$: the coupling constant $g_1=0$ and we have a pure instanton interaction,
where U$_A$(1) is broken maximally.
  \item [(3)]$\alpha=0.5$: in this case $g_1=g_2=G$, the standard NJL model \cite{buballa2}
  is recovered, and only the $\sigma$ and $\pi$ channels, are reproduced.
\end{itemize}

In order to see the effects of the degree of U$_A$(1) symmetry breaking in the vacuum, we
allow $\alpha$ to take values between 0 and 0.5. The results are summarized in Table
\ref{table:results}. As it can be seen, for $\alpha=0$ (no anomaly) the $\eta$ is
degenerated  in mass with the pion, the $a_0$ with $\sigma$, and the topological
susceptibility is zero, as expected. A small breaking of the axial symmetry
$(\alpha=0.1)$ is enough to break the degeneracy, raising the masses of  $\eta$ (that
comes  close  to almost  its experimental value) and $a_0$, and to get  a meaningful
non-vanishing value  of  $ \chi ^{1/4} $. Larger values of $\alpha$ yield larger values
of $\chi^{1/4}$ and of the meson masses. When $0.16\leq\alpha\leq0.25$ the value of
$\chi$ is within the range of values coming from lattice results \cite{Alles,TWQCD}.

\begin{table}[t]
\begin{center}
\begin{tabular}{||cc|cc|cc|c||}
\hline\hline
$\alpha$&& $ \chi ^{1/4} (180)$ & & $\eta(547)$ & & $ a_0(984)$ \\ 
&& [MeV] & & [MeV]             && [MeV]
\\
\hline\hline
0 && 0&&  140.2 && 803.7 \\
0.10 && 150.7 && 510.2&& 856.7\\
0.112 && 155.5 && 547.0&& 863.7\\
0.16 && 170.6 && 641.0&& 893.2\\
0.20 && 180.8 && 704.5&& 919.8\\
0.25 && 191.5 && 766.1&& 956.3\\
0.30 && 200.7 && 799.3&& 996.4\\
0.40 && 216.0 && 986.9&& 1094.3\\
0.50 && 228.6 && --&& --\\
\hline\hline
\end{tabular}\label{dois}
\end{center}
\caption{Numerical values for the topological susceptibility, and masses of the mesons
$\eta$ and $a_0$ for different values of the anomaly parameter, $\alpha$.}
\label{table:results}
\end{table}

\subsubsection{Behavior with temperature}

Let us now discuss our results at finite temperature within the SU(2) PNJL model with
anomaly. We aim at analyzing signals of possible restoration of chiral and axial
symmetries and also to compare these results with those obtained within the corresponding
SU(3) model \cite{Costa:2008PRD,Costa:2009}. First we analyze the characteristic
temperatures for the chiral/deconfinement  phase transition. An important question to
address is whether  U$_A$(1) is effectively restored at finite temperature and its
possible relation with the restoration of chiral symmetry. To quantify this effect it is
mandatory to analyze the behavior of the topological susceptibility, and of  the mesonic
excitations with temperature. Another important question in this concern is to discuss if
$\chi$ should be considered as an order parameter for the axial symmetry. A challenging
question is whether the restoration of axial  symmetry  arises as consequence of the
restoration of chiral symmetry, or if there is an independent mechanism for the
suppression of instanton effects. To simulate the last effect, the anomaly coupling $g_2$
may be assumed as a decreasing function of temperature.

\begin{figure}[t]
\begin{center}
  \begin{tabular}{cc}
  	\hspace*{-0.5cm}\epsfig{file=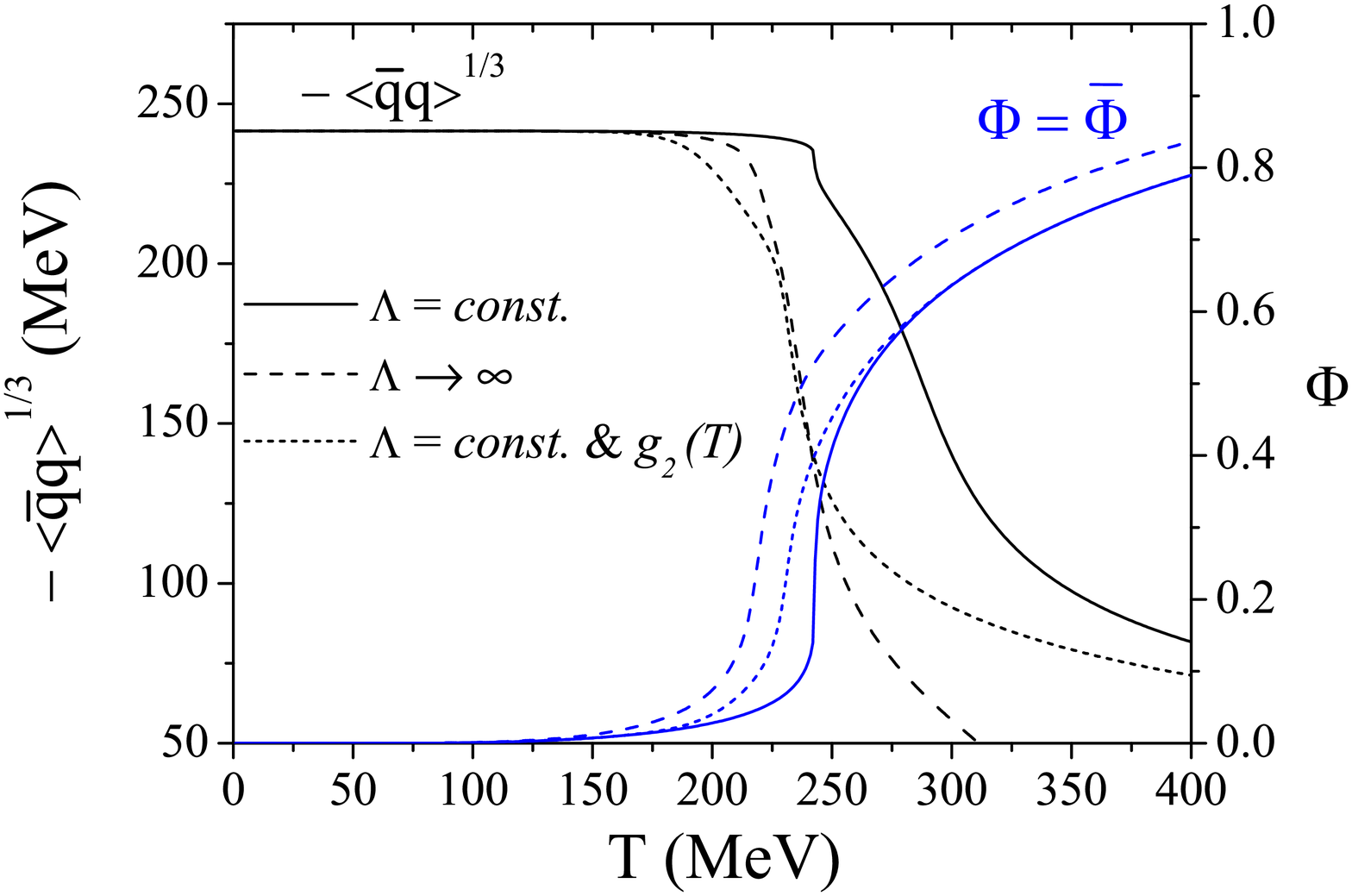,width=8.5cm,height=7.5cm} &
  	\hspace*{-0.5cm}\epsfig{file=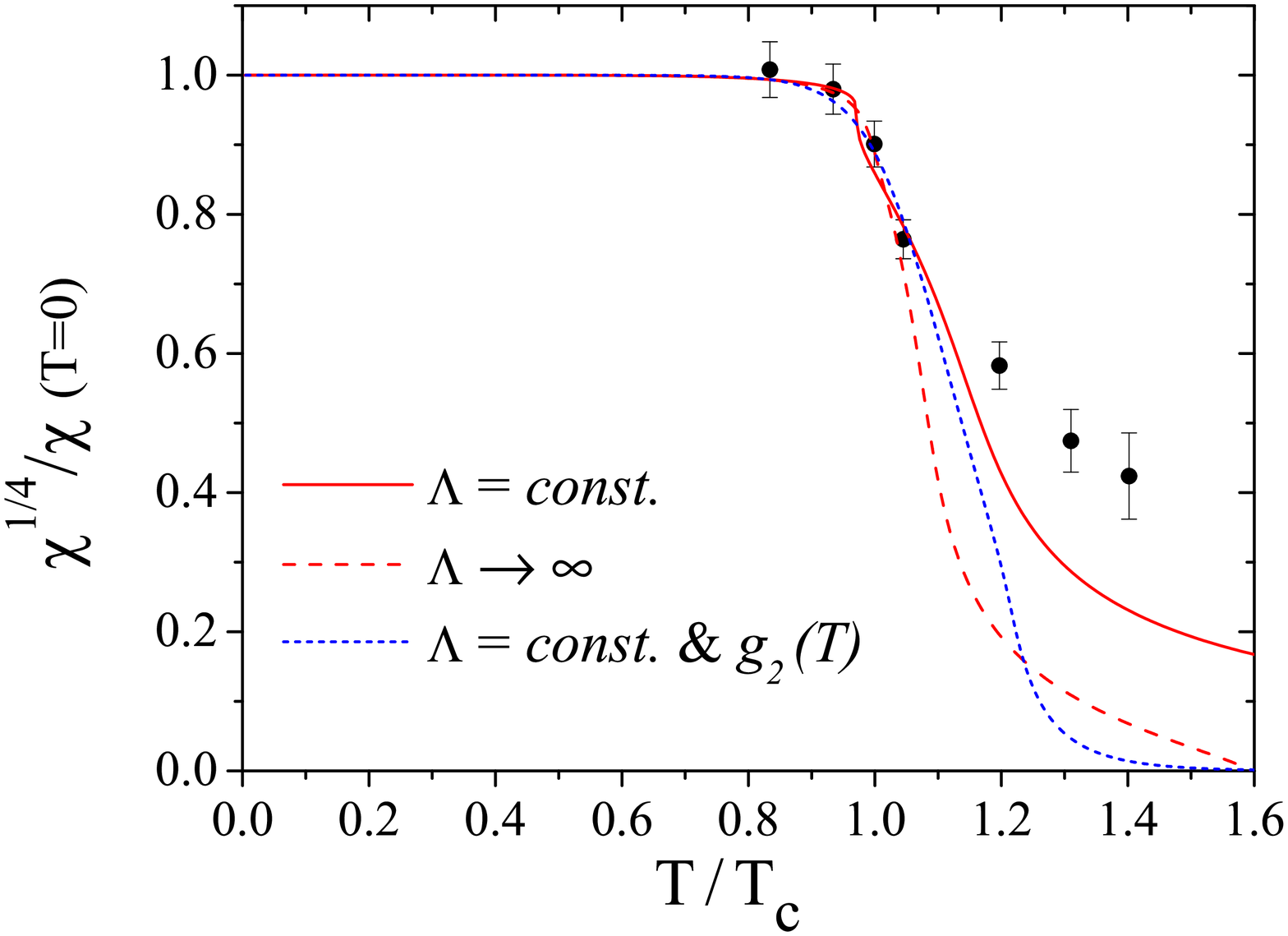,width=8.5cm,height=7.5cm} \\
	\end{tabular}
\end{center}
\caption{Comparison of the behavior with temperature of $-\langle \bar qq\rangle^{1/3}$
and the Polyakov field $\Phi$ (left panel) and  of the topological susceptibility (right
panel) with $g_2$ constant and two regularizations ({\it I}: $\Lambda \rightarrow
\infty$, {\it II}: $\Lambda\,=\,const.$) and with $g_2$ a decreasing function of
temperature and \textit{regularization II} ($\alpha=0.2$). The lattice points for $\chi$
were taken from Ref. \cite{Alles}.}
\label{Fig:cond}
\end{figure}

We start by discussing the behavior with temperature of observables that may signal
restoration of chiral symmetry, deconfinement and restoration of axial symmetry in
different situations. For this purpose, let us first analyze in Fig. \ref{Fig:cond} (left
panel) the behavior of the quark condensate $-\langle \bar qq\rangle^{1/3}$, of  the
Polyakov field $\Phi$ (left panel) and  of the topological susceptibility (right panel)
with $g_2$ constant and two {\it regularizations} ({\it I}: $\Lambda \rightarrow \infty$,
{\it II}: $\Lambda\, =\, const.$). From now on we will focus our discussion for the case
$\alpha=0.2$ (the analysis is qualitatively similar for others values of $\alpha$.)

The results for  the quark condensate and for the Polyakov loop are similar to those
within the SU(3) model. As it was shown in \cite{sym,ruivo}, the critical temperature for
the chiral and deconfinement transitions, identified with the derivatives of  the quark
condensate and of the Polyakov field, respectively, become closer using
\textit{regularization I}. As shown in \cite{Ratti:2005PRD,Costa:2008PRD} this
regularization, together with the Polyakov loop, is very convenient to describe the
thermodynamic properties of the system leading to a good agreement with lattice results.
It has the disadvantage, for high temperatures, of leading to a sharp decrease of the
quark condensate that, when using \textit{regularization I}, vanishes and, unless an
additional condition is implemented, becomes negative.

In order to discuss the influence of an independent mechanism of suppression of
instantons with temperature, we consider a third situation: the anomaly coupling is
assumed to decrease with $T$, $g_2(T) = g_2(0)e^{-(T-300)/20}$, and
\textit{regularization II} is used. Since {\it Scenario A} is considered, the variation
of $g_2$ with $T$ implies also a variation of $g_s= 2(g_1 + g_2)$, and therefore the
``mechanism of instanton suppression''  influences the restoration of chiral symmetry and
deconfinement. In fact, we can see from Fig. \ref{Fig:cond} that both $-\langle \bar
qq\rangle^{1/3}$ and $\Phi$ are affected by the decrease of $g_2$. By examining the
values of Table \ref{table:temps} we see that, concerning the critical temperatures, both
this case and \textit{regularization I} with $g_2$ constant lead to a greater proximity
of the deconfinement and chiral phase transitions as compared with \textit{regularization
II} and $g_2$ constant. We emphasize that the ansatz $g_2 (T)$ leads even to equal values
for $T_c^\chi$ and $T_{c}^{\Phi}$.


\begin{table}[t]

\begin{center}

\begin{tabular}{|| l|| c| c| c||}

\hline\hline       & \textit{regularization I}         & \textit{regularization II}    &
\textit{regularization II}  \\ [-3mm]
         & ($\Lambda \rightarrow \infty$)  & ($\Lambda\, =\, const.$)            &  with $g_2 (T)$ \\
\hline\hline

$T_c^\chi$ [MeV]      & 237                  & 287                                           & 231\\

\hline

$T_{c}^{\Phi}$ [MeV]  & 219                & 243                                           & 231\\

\hline

$T_{c}$ [MeV]         & 228              & 265                                           & 231\\

\hline

$T_{eff}^\chi$ [MeV]  & $\simeq$ 260              & $\simeq$ 350                              & $\simeq$ 245\\

\hline

$T_{eff}^A$ [MeV]     & ---                 &  ---                                          & $\simeq$ 285\\

\hline\hline

\end{tabular}

\caption{Critical temperatures for chiral and deconfinement phase transitions ($T_c^\chi$ and $T_{c}^{\Phi}$;
$T_{c}$ is the average between both) and for the effective restoration of chiral and axial symmetries
($T_{eff}^\chi$ and $T_{eff}^A$ ), obtained   for $\alpha = 0.2$.
The values in the two first rows were obtained with finite and infinite cutoff, respectively, and $g_2$ constant;
in the third row $g_2$ is a decreasing function of temperature and the cutoff is kept constant.}
\label{table:temps}

\end{center}

\end{table}

Observing the right panel of Fig. \ref{Fig:cond}, we see that, as already shown in the
SU(3) model \cite{Costa:2009}, PNJL calculation for the topological susceptibility nicely
reproduces the first lattice points, a feature that is   not verified in the NJL model
\cite{Costa:2009}. The effects of the introduction of high quark moments through
\textit{regularization I} is only relevant at high temperatures, allowing a decrease of
$\chi$, that eventually becomes zero. The vanishing of $\chi$ is also achieved with a
decreasing with temperature of the strength of the anomaly ($g_2(T)$). So, after
analyzing both figures, it seems that two independent mechanisms, the allowance of hight
momentum quark states  and the instanton suppression, produce some similar effects:
proximity between chiral and deconfinement phase transitions, and sharp decrease and
vanishing of the topological susceptibility. The question is whether the mentioned
behavior of the topological susceptibility is enough to signal restoration of the axial
symmetry.

With rising temperature we expect the chiral
SU(2)$_V\otimes$ SU(2)$_A\,\simeq \,$SU(2)$_L\otimes$SU(2)$_R$
symmetry to be restored, and the chiral partners
($\pi,\,\sigma$) and ($\eta,\, a_0$) become degenerate in mass, the effective restoration of chiral
symmetry being then achieved.

Concerning the axial symmetry, the vanishing of the mass splitting of the axial chiral
partners ($\pi, a_0$) and ($\sigma, \eta$)  is a necessary condition for the restoration
of this symmetry. In order to discuss this problem, we are going to analyze the behavior
with temperature of the topological susceptibility and meson masses by exploring two
situations:  (i)  using the two regularizations already presented, but without allowing
the anomaly coupling $g_2$ to decrease with temperature, (ii)  considering several
degrees of violation of axial symmetry \textit{ab initio} by using three different values
for the anomalous parameter $\alpha$. Once again we will consider a fixed $\alpha=0.2$
for $T=0$.

\begin{figure}[t]
\begin{center}
  \begin{tabular}{cc}
    \hspace*{-0.5cm}\epsfig{file=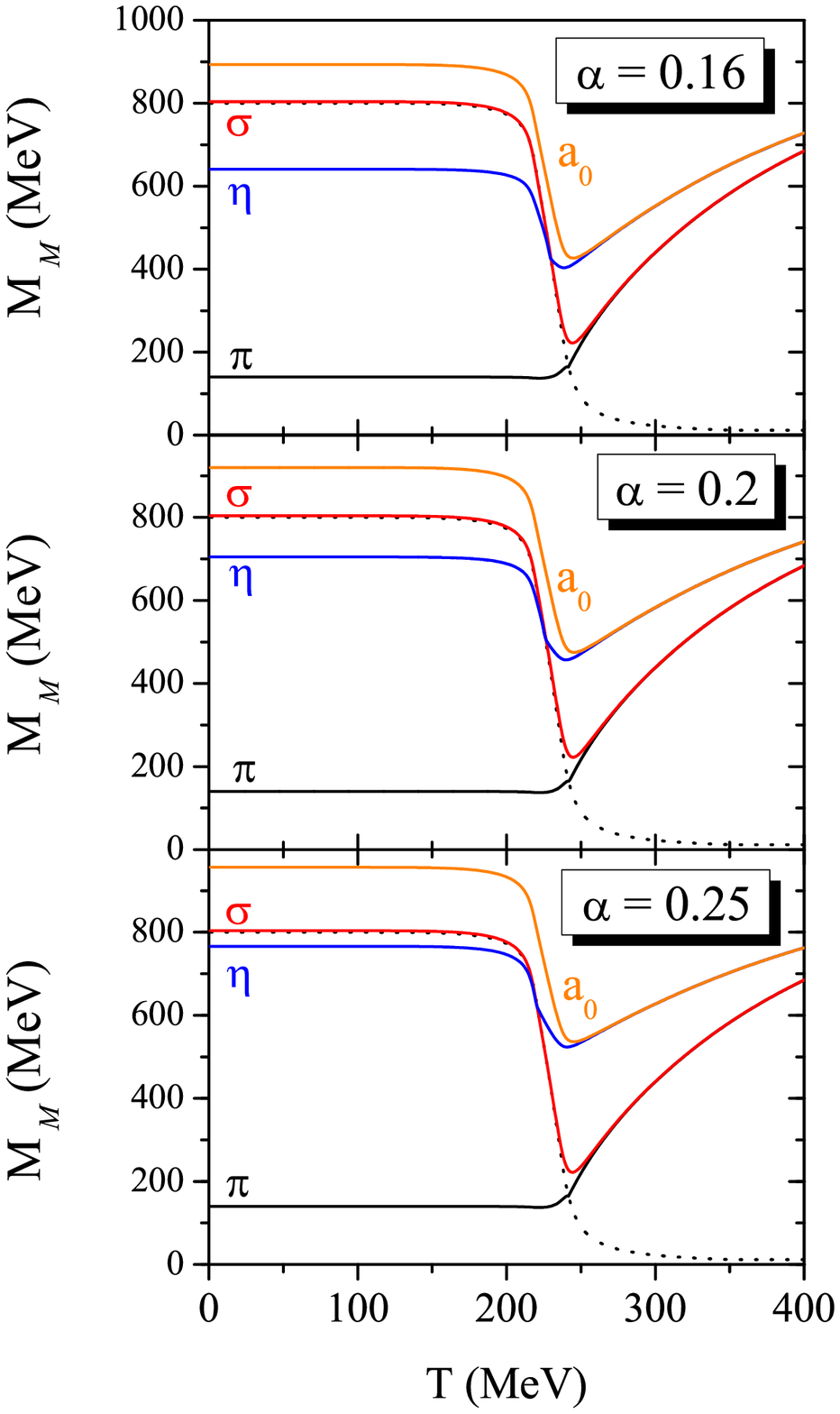,width=8.5cm,height=12cm} &
    \hspace*{-0.5cm}\epsfig{file=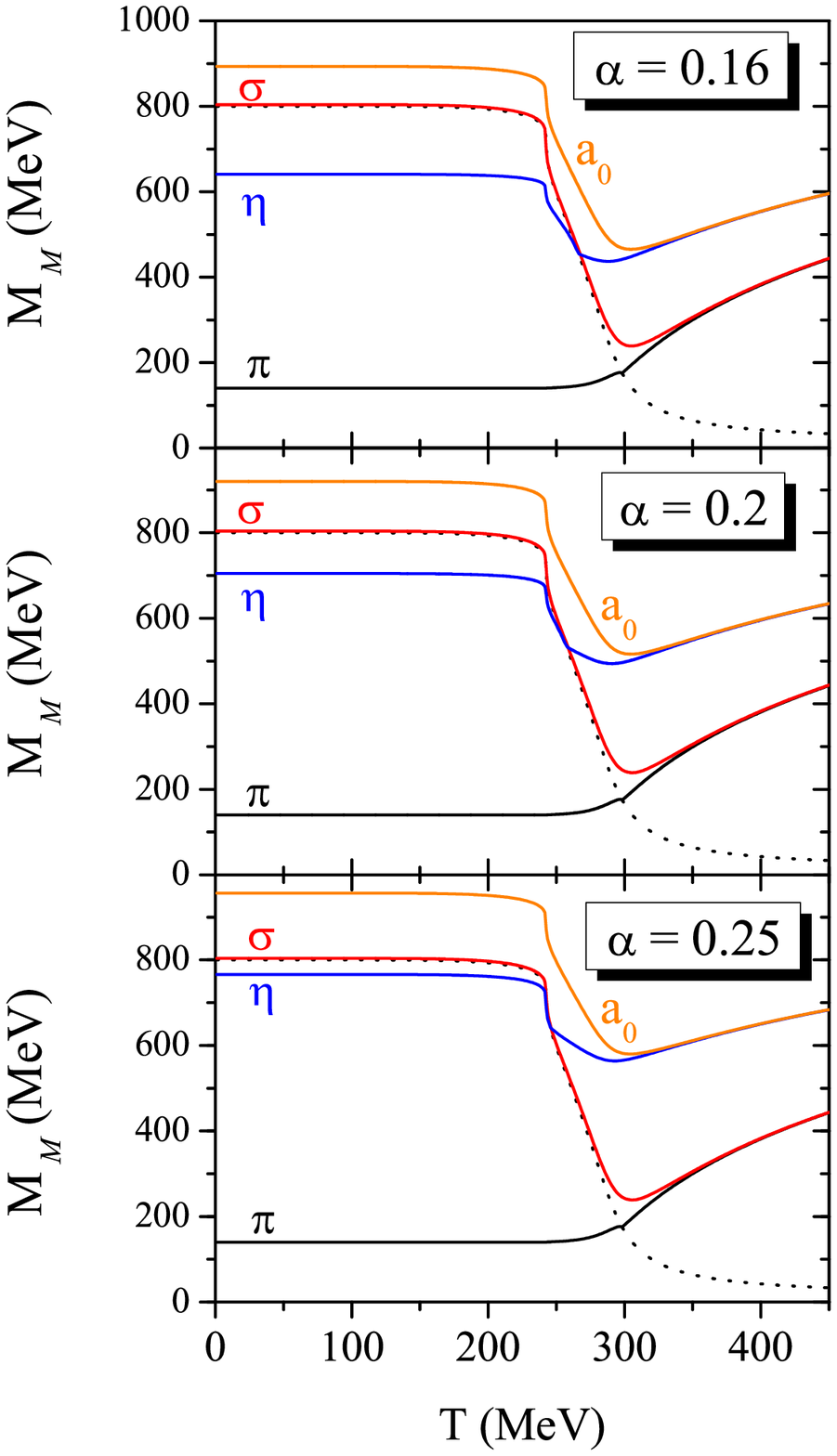,width=8.5cm,height=12cm}
  \end{tabular}
\end{center}
\label{mesons} \caption{Masses of the meson chiral partners  $(\pi, \sigma)$ and $(\eta,
a_0)$ with varying temperature for three different values of the anomaly parameter
$\alpha$ with two regularizations at finite $T$: \textit{regularization I} (left panel)
and \textit{regularization II} (right panel). The dotted lines represent the $q\bar q$
threshold  $2 M_u$.} \label{Fig:mesoes1}
\end{figure}

\begin{figure}[t]
\begin{center}
\begin{tabular}{cc}
    \hspace*{-0.5cm}\epsfig{file=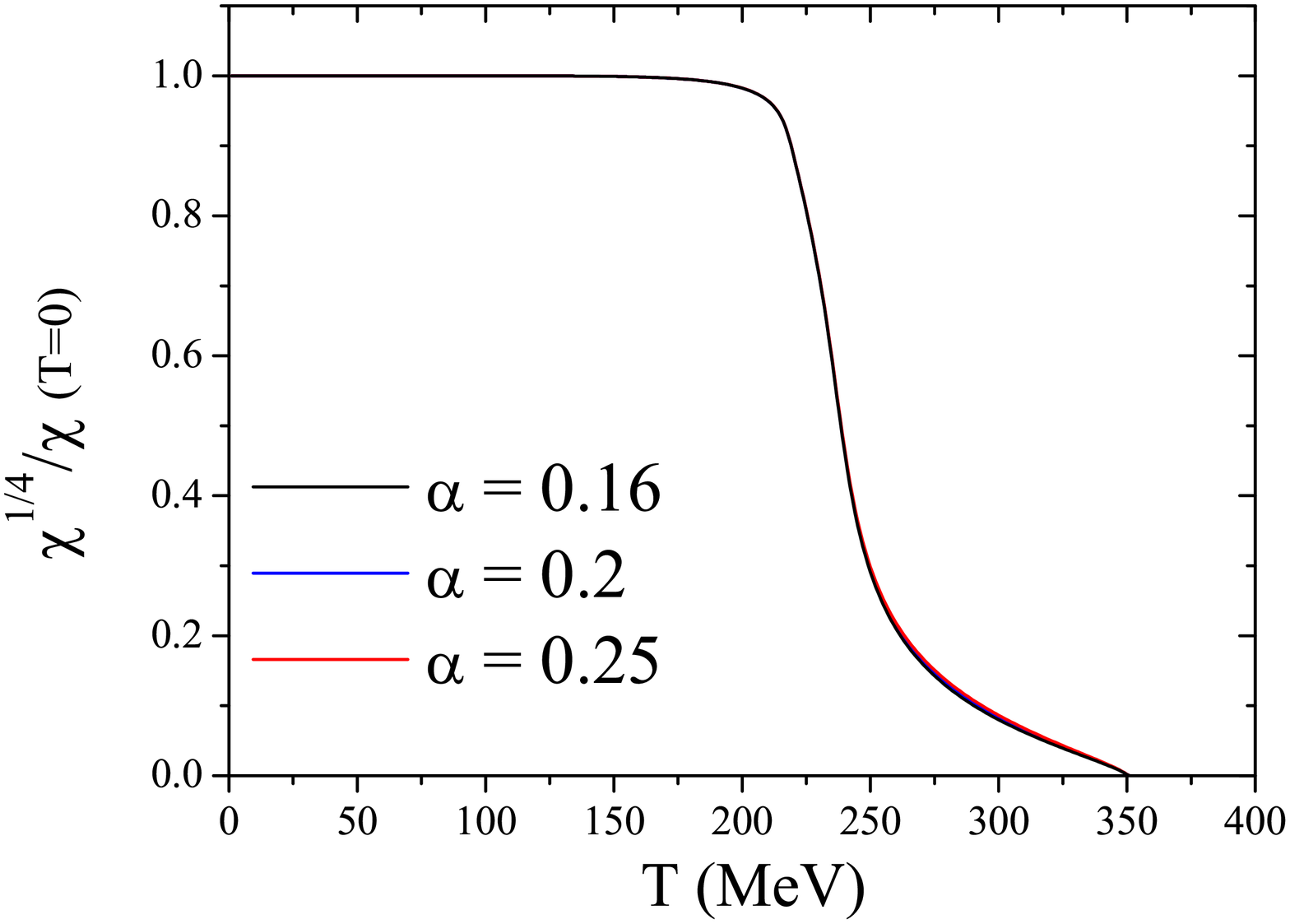,width=8.5cm,height=7.cm} &
    \hspace*{-0.5cm}\epsfig{file=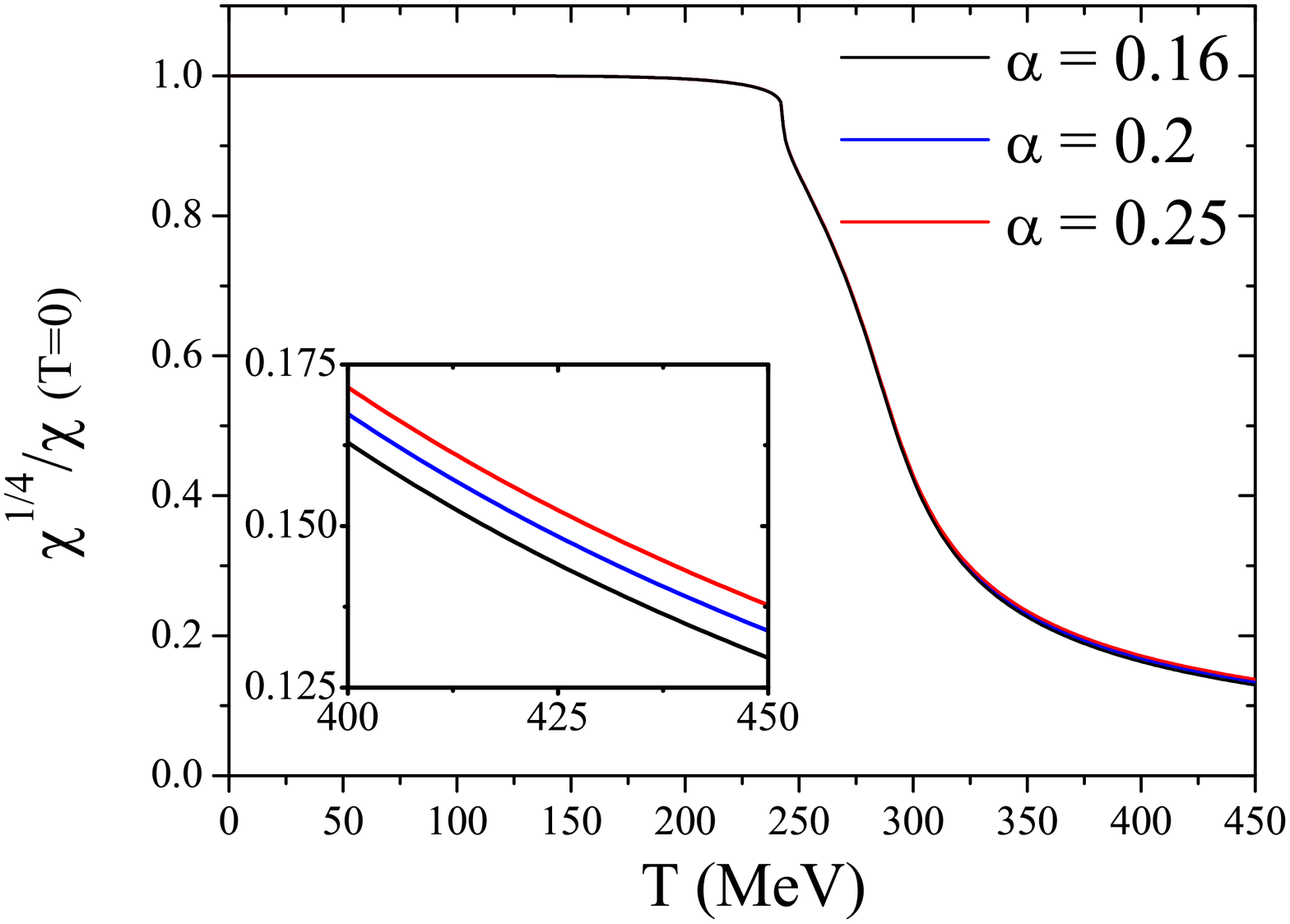,width=8.5cm,height=7.cm}\\
\end{tabular}
\end{center}
\caption{Normalized topological susceptibility as function of temperature for several
values of the anomaly parameter $\alpha$ with {\it regularizations I} (left panel) and
{\it II} (right panel).}
\label{Fig:susc1}
\end{figure}

The temperature dependence of the mesonic observables is displayed in Fig.
\ref{Fig:mesoes1}. Concerning the chiral partners ($\sigma,\,\pi$), we observe that in
the vicinity of 260 (350) MeV for \textit{regularization I} (\textit{regularization II})
both masses become identical, a sign for the effective restoration of chiral symmetry. At
a temperature slightly lower there occurs the convergence of the chiral partners ($\eta,
a_0$). The interesting result is that the restoration of the axial symmetry is not
achieved: the masses of the two partners ($\pi,\,\sigma$) and ($\eta,\, a_0$), although
getting close at high temperatures, do not converge.  Concerning the variation of
$\alpha$,  the lower values of this parameter favor the restoration of axial symmetry, as
expected,  since the masses of the axial chiral mesons were already closer in the vacuum.
Even when high momentum quarks are allowed (left panel) the masses of partners, although
closer, do not converge (see Fig. \ref{Fig:mesoes1}, right panel), in spite of the
vanishing of the topological susceptibility in this case (see Fig. \ref{Fig:susc1}).

This result is different from the one obtained in the SU(3) versions of NJL and PNJL
models \cite{Costa:2009}, where one could obtain a simultaneous vanishing   of the
splitting of the axial partners and of the topological susceptibility when
\textit{regularization I} was used. In fact, the inclusion of  strange quarks (N$_f$ = 3)
leads to  non-linear (mixing) effects since the U$_A$(1) anomaly term is trilinear for
three flavors and thus would generate additional mechanisms of restoration of axial
symmetry, even  if a constant anomaly strength is taken into account. In order to
understand the present results, let us remember the comparison of the expressions for
$\chi$ within the SU(3) and SU(2) models as discussed in Sec. \ref{subsec:topsuc}.

A meaningful difference between both models is that in SU(3) we can define an effective
anomaly strength, $g_{D_{eff}}= g_D  \left\langle \bar q q \right\rangle $, that, as the
temperature varies, acquires a temperature dependence through the variation with
temperature of the quark condensate. Therefore it is possible to obtain restoration of
chiral axial symmetry without additional assumptions, the infinite cutoff at finite $T$
leads to the vanishing of the quark condensate at high temperatures and, as a consequence
of the effective anomaly strength and of the quantities that depends on it: the
topological susceptibility and the mass splitting  between the axial partners. In this
case restoration of axial symmetry is a mere consequence of the full restoration of
chiral symmetry. It is not necessary to implement an anzats related to explicit instanton
suppression. As we will  show, the vanishing of the mass splitting between axial partners
in SU(2) demands such a mechanism.

\begin{figure}[t]
\begin{center}
\begin{tabular}{cc}
    \hspace*{-0.5cm}\epsfig{file=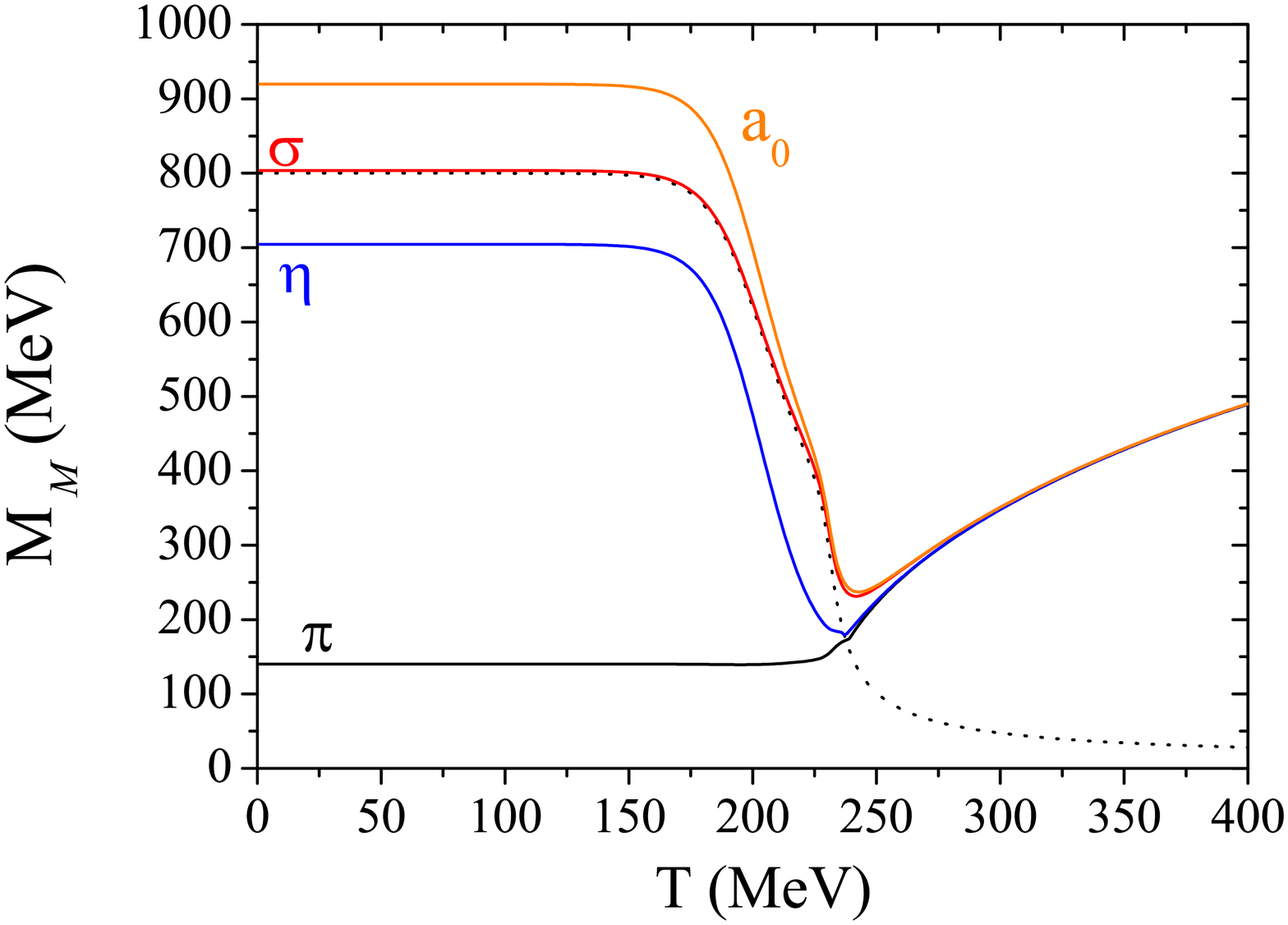,width=8.5cm,height=7.cm} &
    \hspace*{-0.5cm}\epsfig{file=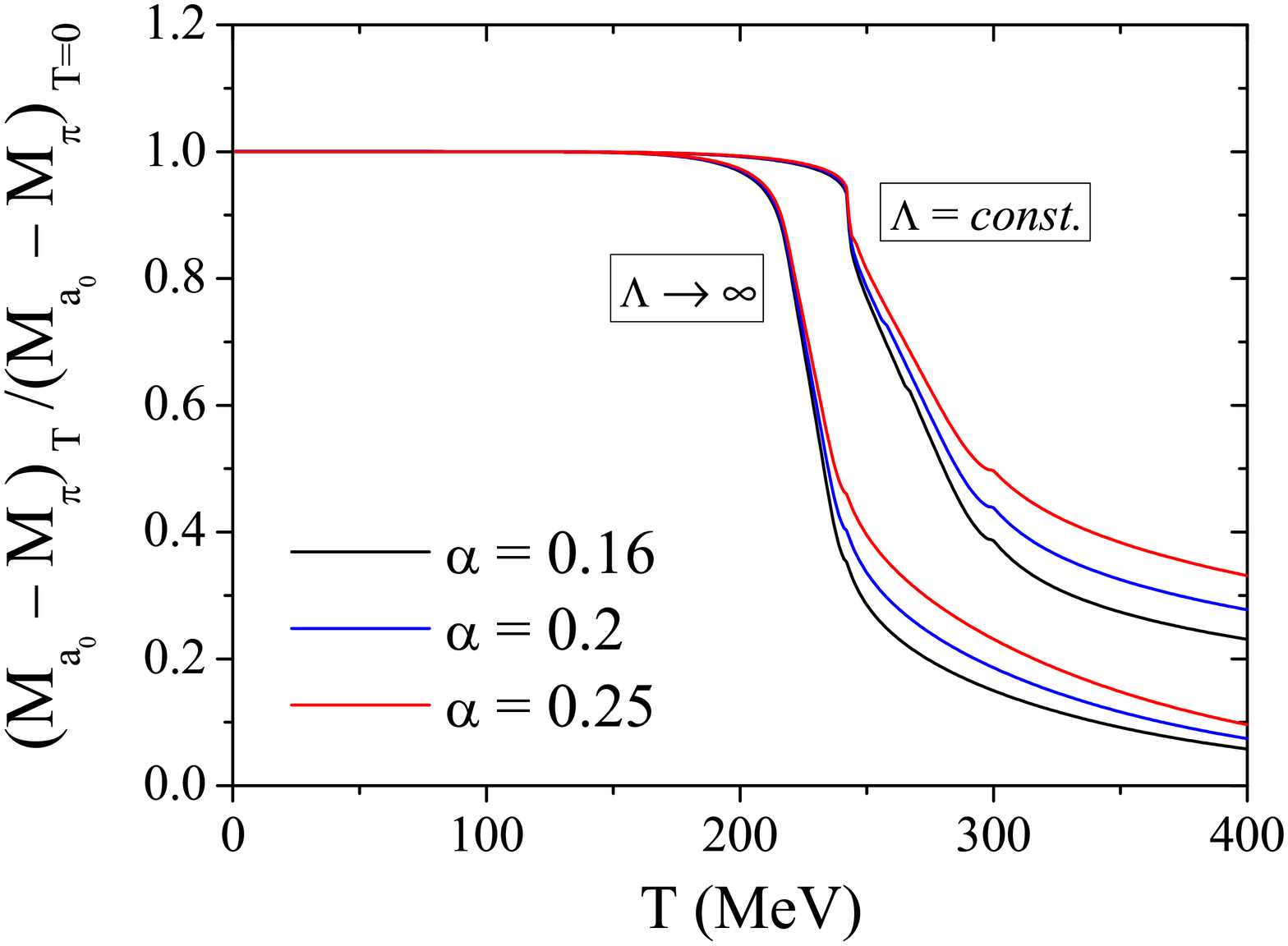,width=8.5cm,height=7.cm}\\
\end{tabular}
\end{center}
\caption{Left panel: masses of the meson chiral partners  $(\pi, \sigma)$ and $(\eta,
a_0)$ with varying temperature for $g_2(T)$  as a decreasing function of temperature,
using \textit{regularization II} ($\Lambda\,=\,const.$) at  finite $T$.  The dotted line
represents the $q\bar q$ threshold  $2 M_u$. Right panel: normalized mass splitting of
axial partners as function of temperature for different values of $\alpha$.}
\label{Fig:mesoes2}
\end{figure}

The results plotted in Figs. \ref{Fig:cond}--\ref{Fig:susc1} already give an indication that a weaker
axial symmetry breaking (in this case obtained by hand, by giving low values to $\alpha$) favors the
restoration of axial symmetry and reduces the mass splitting of the axial partners (see
Fig. \ref{Fig:mesoes1}). In fact, the restoration of this symmetry can only be modeled in a
phenomenological way by making $g_2$ temperature dependent. Therefore we will  calculate
the meson mass spectrum with $g_1$ fixed and $g_2$  a decreasing exponential of the
temperature, $g_2(T) = g_2(0)e^{-(T-300)/20}$. The result is plotted in Fig. \ref{Fig:mesoes2}
where we can see that, although the effective restoration of chiral symmetry (vanishing of the
mass splitting between $\sigma$ and $\pi$, and between $\eta$ and $a_0$) occurs first, at
$T=245$ MeV, the degeneracy of the $\pi, \,\eta$, $a_0$ and  $\sigma$ occurs at $T=285$
MeV. Therefore, although the analysis of the effects of high momentum quark states and of
the instanton suppression have similarities in what concerns the characteristic
temperatures for the phase transitions and the high temperature behavior of the
topological susceptibility, the evidence that they are indeed mechanisms with a different
physical meaning appears when we look for a possible degeneracy of the axial chiral
partners.

\section{Conclusions}
\label{sec:conclusions}

The present study is dedicated to the analysis of the  behavior of various observables
that signal the restoration of the chiral and axial symmetries in regards to temperature,
within a SU(2) Polyakov--Nambu--Jona-Lasinio model with a 't Hooft interaction term.

Two types of regularization of finite temperature were considered; the first
regularization consists in using the cutoff only on divergent integrals ({\it I}) and the
other consists in always using a finite cutoff ({\it II}). The effect of the anomalous
coefficient dependent on temperature, $g_2(T)$, was also analyzed.
The restoration of chiral and axial symmetries with temperature was studied, considering
the effects in the symmetry restoration process of the type of regularization applied. A
comparative study  between the results here obtained and the ones calculated within the
PNJL model in SU(3)~\cite{ruivo} was also performed. Similarly to SU(3), it  was verified
in the present work that the critical temperatures  obtained with {\it regularization I}
are closer to lattice results than with  {\it regularization II}; however, the effective
restoration of the chiral and axial symmetries does not occur simultaneously.

We hence conclude that when the coupling constants are considered constant, the
restoration of the axial symmetry, in the SU(2) sector, does not occur as a natural
consequence of the complete restoration of the chiral symmetry, being this fact a
fundamental difference between the SU(3)  and the SU(2) sectors, where it is not
sufficient to apply the {\it regularization I} for all the observables, related with the
U$_A$(1) symmetry, to vanish. It is therefore necessary to use an additional mechanism
for this restoration, such as  a  dependence on the temperature of the coupling
coefficient that is related to the anomaly. Once $g_2$ is the coupling constant of the
term that breaks the axial symmetry and simulates the instanton effect, this result seems
to indicate that, in SU(2), the instantons are not completely suppressed by the
restoration of the chiral symmetry.

A relevant contribution that the present study offers for the understanding of physics
associated with the breaking and restoration of the U$_A$(1) symmetry refers to the fact
that the analysis of the topological susceptibility, $\chi$, is the ``necessary condition'',
but not the ``sufficient condition'', for the study and comprehension of the restoration
of the axial symmetry.

\begin{acknowledgments}
Work supported by  Centro de F\'{\i}sica Computacional and F.C.T.
under Project No. CERN/FP/116356/2010.
\end{acknowledgments}

\vspace{0.5cm}

\appendix
\section{}
\label{appen}

Some meaningful differences between the results
within the SU(3) and the SU(2) models appear, so it is useful, in order to understand
these differences, to present the  basic ingredients of  the SU(3)  model. The SU(3) NJL
model with  the 't Hooft determinant has the following Lagrangian:
\begin{eqnarray}
{\cal L} &=& \bar{q} \left( i \gamma^\mu \partial_\mu - \hat{m} \right) q +
\frac{g_S}{2} \sum_{a=0}^{8} \Bigl[ \left( \bar{q} \lambda^a q \right)^2+
\left( \bar{q}\, i \gamma_5 \lambda^a\, q \right)^2 \Bigr] \nonumber \\
&+& g_D \Bigl[ \mbox{det}\bigl[ \bar{q} (1+\gamma_5) q \bigr] + \mbox{det}\bigl[ \bar{q}
(1-\gamma_5) q \bigr]\Bigr] \, ,\label{lagr}
\end{eqnarray}
which can be rewritten as:
\begin{eqnarray}\label{NJL}
  {\cal L} =
\bar{q} \left( i \gamma^\mu \partial_\mu - \hat{m} \right) q + \frac{1}{2} \Bigl\{ (\bar{q}
\lambda^a q)S_{ab}(\bar{q} \lambda^b q) +(\bar{q}\, i\gamma_5 \lambda^a\, q) P_{ab}
(\bar{q}\, i\gamma_5 \lambda^b\, q) \Bigr\} \, ,
\end{eqnarray}
where the following projectors have been introduced:
\begin{eqnarray}
  S_{ab} &=& g_S \delta_{ab} + g_D D_{abc}\left\langle \bar{q} \lambda^c q\right\rangle \, , \label{sab}\\
  P_{ab} &=& g_S \delta_{ab} - g_D D_{abc}\left\langle \bar{q} \lambda^c q\right\rangle \, . \label{pab}
\end{eqnarray}
 $\left\langle \bar{q} \lambda^c q\right\rangle$ being the  vacuum expectation values. The constants $D_{abc}$
coincide with the SU(3) structure constants $d_{abc}$ for a, b, c = (1, 2, . . . , 8) and
$D_{0bc} = - \frac{1}{\sqrt{6}} \delta_{bc}$ and  $D_{000}= \sqrt{\frac {2}{3}}$.


\end{document}